\documentclass[aps,prb,twocolumn]{revtex4}
\usepackage{graphicx}
\usepackage{dcolumn}
\usepackage{amsmath}
\usepackage{amssymb}
\def\uu{\uparrow\uparrow}
\def\ud{\uparrow\downarrow}
\def\du{\downarrow\uparrow}
\def\dd{\downarrow\downarrow}
\def\rv{{\bf r}}
\def\Rv{{\bf R}}
\def\kv{{\bf k}}
\def\Kv{{\bf K}}
\def\e{{\rm e}}

\begin{document}
\title{Short-range correlation in the uniform electron gas: Extended
Overhauser model}
\author{Paola Gori-Giorgi$^{1,2}$\footnote{present address:
Dipartimento di Fisica, Universit\`a di Roma ``La Sapienza'', 
Piazzale A. Moro 2, 00185 Rome, Italy} and 
John P. Perdew$^1$}
\affiliation{$^1$Department of Physics and Quantum Theory Group, Tulane
University, New Orleans, Louisiana 70118 USA\\
$^2$Unit\`a INFM di Roma ``La Sapienza'', Piazzale A. Moro 2,
00185 Rome, Italy}
\date{\today}
\begin{abstract}
We use the two-electron wavefunctions (geminals) and
the simple screened Coulomb potential proposed by Overhauser
[Can. J. Phys. {\bf 73}, 683 (1995)] to compute the
pair-distribution function $g(r)$ for a uniform electron gas,
finding the exact $g(0)$ for this model and extending the
results from $g(0)$ to $g(r)$. We find that
the short-range ($r<r_s$) part of this $g(r)$ is in excellent agreement 
with Quantum Monte
Carlo simulations for a wide range of electron densities. We are thus
able to estimate the value of the second-order ($r^2$) coefficient
of the small interelectronic-distance expansion of the pair-distribution 
function. 
The coefficients of the small-$r$ expansion of the spin-resolved
$g_{\sigma\sigma'}(r)$ have density or $r_s$ dependences which 
we parametrize in a
way that makes it easy to find their coupling-constant averages.  Their
spin-polarization or $\zeta$ dependences are estimated from a 
proposed spin scaling relation.
\end{abstract}
\pacs{71.10.Ca, 71.10.-w, 05.30.Fk, 71.15.Mb}
\maketitle
\section{Introduction}
The pair-distribution function $g(r)$ for the 
uniform electron gas is related to the probability of finding a
pair of electrons at a distance $r$ from each other.
(The expected number of electrons in a shell
of volume $4\pi r^2dr$ when another electron is at the origin
is given by $ng(r)4\pi r^2 dr$, where $n=N/V$ is 
the electron density.)
In recent years, much attention has been devoted to this
quantity,\cite{Kimball2,Sacchetti,PW,Rassolov,GSB,comment,reply}
mainly because of its relevance in density-functional
theory: $g(r)$ is the basis of many non-local\cite{WDA,Chacon}
and semi-local\cite{GGA} exchange-correlation energy density functionals.
A good $g(r)$ for the uniform electron gas is also the starting
point for building up the system-averaged
exchange-correlation hole of a many-electron
system of non-uniform density.\cite{nxc}

After oscillations are averaged out,
the long-range part ($r\to \infty$) of $g(r)$ is exactly 
described\cite{NP58,PW} by the random-phase approximation (RPA). 
At intermediate interelectronic distances, $0.5\lesssim r/r_s\lesssim 5$ 
(where $r_s$
is the density parameter, $r_s=(4\pi\,n/3)^{-1/3}$), very reliable
information is available from Quantum Monte Carlo (QMC) 
simulations.\cite{cepald,Pickett,OB,OHB}
Little is known
about the quantitative behavior of $g(r)$ for $r/r_s\lesssim 0.5$,
except in the high-density limit,\cite{Rassolov,reply} and except
some estimates of $g(0)$, the value of the pair-distribution
function at zero-interelectronic
distance.\cite{PW,GSB,Yasuhara,Ov,BPE,Geldart} 
(As the part of $g(r)$ that is most
transferable from uniform to non-uniform densities, $g(0)$ plays a
special role in spin-density functional theory.\cite{BPE})

In particular, Overhauser~\cite{Ov} presented a simple model for estimating
the value of $g(0)$. The model takes into account
two-particle correlations by means of a simple screened Coulomb potential
with no empirical parameter. In his work Overhauser finds an approximate
solution for the dependence of $g(0)$ on the electron density. 
His result looks realistic and performs
surprisingly well in the known high-density limit.

Motivated by this result,
in the present work we solve the Overhauser model exactly, i.e., we compute
the corresponding pair-distribution function 
for the uniform electron gas. Comparison of
our results with recent QMC 
simulations~\cite{OHB} shows very good agreement
in the range $0.5 \lesssim r/r_s \lesssim 1$. We thus have a strong
indication that the Overhauser model potential gives accurate
quantitative results also in the ``unknown'' shortest-range 
region $0 \le r/r_s \lesssim 0.5$. In this way we are able to
present a quantitative, reliable
estimate for the $r_s$ dependence of the $r^2$ coefficient of the
small-$r$ expansion of $g(r)$. This coefficient is 
important for new energy density functionals which include the 
gradient correction to the $r^2$ coefficient of the 
exchange-correlation hole, while its spin resolution (i.e.,
its $\uu$ and $\ud$ contributions, also available
in the present treatment) is of interest for
functionals based on the Fermi hole curvature.\cite{Dobson} Other
possible applications are discussed at the end of this article.

The Overhauser model is fully quantum mechanical, and incorporates
the effect of exchange on the pair-distribution function in the 
zero-temperature electron gas. An interesting alternative 
approach\cite{mimT} mimics these effects by using a classical
pair-distribution function with a non-zero effective temperature. 

\section{Uniform electron gas from two-particle wavefunctions}

Following Overhauser,\cite{Ov} we will construct the pair distribution
function from two-electron wavefunctions.  The rigorous analogs of these 
wavefunctions are perhaps the ``natural geminals'' that diagonalize the
two-electron density matrix;\cite{D,Z} the diagonal of this matrix is the pair
density. For a generalization of the Hartree-Fock
approximation from orbitals to geminals, see Ref.~\cite{GS}.

If we select a pair of electrons at random in the spin-unpolarized 
uniform gas, there is one chance in four that they will be in the singlet
state, $\ud-\du$, and three chances in four that they will be in one of the
triplet states, $\uu$, $\dd$, $\ud+\du$. In the case of no 
electron-electron interaction, the corresponding two-electron 
spatial wavefunctions will be
\begin{equation}
\Psi(\rv,\Rv)=\frac{1}{\sqrt{2}}\,\e^{i\,\Kv\cdot\Rv}\,\left(
\e^{i\,\kv\cdot\rv}\pm\e^{-i\,\kv\cdot\rv}\right),
\label{eq_freewave}
\end{equation}
where ``$+$''is for the singlet state and ``$-$'' is for
the triplet state, and
\begin{eqnarray*}
 \Rv=\tfrac{1}{2}(\rv_1+\rv_2), \qquad  \rv=\rv_2-\rv_1, \\
 \Kv=\kv_1+\kv_2, \qquad \kv=\tfrac{1}{2}(\kv_2-\kv_1).
\end{eqnarray*}
We can expand the plane waves into spherical harmonics
\begin{equation}
\e^{i\,\kv\cdot\rv}=\sum_{\ell=0}^{\infty}(2\ell+1)\,i^{\ell}\,P_{\ell}(\cos
\theta)\,j_{\ell}(k\,r),
\end{equation}
where $P_{\ell}$ are Legendre polynomials and $j_{\ell}$ are spherical Bessel
functions. Then we will have
\begin{eqnarray*}
\Psi_{\mathrm{singlet}}(\rv,\Rv) & = & \sqrt{2}\,\e^{i\,\Kv\cdot\Rv}
\sum_{\stackrel{\ell=0}{\mathrm{even}\; \ell}}^{\infty}(2\ell+1)\,i^{\ell}\,
P_{\ell}(\cos\theta)\,j_{\ell}(k\,r) \\
\Psi_{\mathrm{triplet}}(\rv,\Rv) & = & \sqrt{2}\,\e^{i\,\Kv\cdot\Rv}
\sum_{\stackrel{\ell=1}{\mathrm{odd}\; \ell}}^{\infty}(2\ell+1)\,i^{\ell}\,
P_{\ell}(\cos\theta)\,j_{\ell}(k\,r).
\end{eqnarray*}
We can define spin-resolved pair-distribution functions for the uniform
electron gas, $g_{\uu}(r)$ and $g_{\ud}(r)$, corresponding to parallel- and
antiparallel-spin interactions, and such that for the unpolarized
gas 
\begin{equation}
g=\tfrac{1}{2}(g_{\uu}+g_{\ud}). 
\end{equation}
They can be related to the formulas
just derived by simple considerations: (i) if we select at random a pair 
of electrons in the uniform gas, there is a
probability $p(k)$ that they have relative momentum $k=|\kv_2-\kv_1|/2$,
(ii) the probability for
the singlet state is $\frac{1}{4}$ and the probability for the triplet
 state is $\frac{3}{4}$,
(iii) $\frac{1}{3}$ of the triplet state contributes to the antiparallel-spin 
correlations and $\frac{2}{3}$ of it to the parallel-spin correlations. 
So, we simply have
\begin{eqnarray}
g_{\ud}(r) & = & \tfrac{1}{2}\langle |\Psi_{\mathrm{singlet}}(\rv)|^2 \rangle
+\tfrac{1}{2}\langle |\Psi_{\mathrm{triplet}}(\rv)|^2 \rangle\\
g_{\uu}(r) & = & \langle |\Psi_{\mathrm{triplet}}(\rv)|^2\rangle,
\end{eqnarray}
where $\langle \rangle$ denotes average over $p(k)$ and over the solid angle.
Performing the spherical average over the solid angle, we obtain:
\begin{eqnarray}
g_{\ud}(r) & = & \sum_{\ell=0}^{\infty}(2\ell+1)\langle 
j_{\ell}^2(k\,r)\rangle
\label{eq_gudnonint} \\
g_{\uu}(r) & = & 2\sum_{\stackrel{\ell=1}{\mathrm{odd}\; \ell}}^{\infty}
(2\ell+1)\langle j_{\ell}^2(k\,r)\rangle.
\label{eq_guunonint}
\end{eqnarray}
Eq.~(\ref{eq_gudnonint}) immediately gives the exact result for a 
noninteracting gas, i.e., $g_{\ud}(r)=1$ for each $r$. To obtain the 
noninteracting $g_{\uu}(r)$ from Eq.~(\ref{eq_guunonint}), we need
to average over $k$. In the noninteracting electron gas,
the probability distribution $p(k)$ for 
$k=\tfrac{1}{2}|\kv_2-\kv_1|$ can
be obtained geometrically by considering two three-dimensional 
vectors $\kv_1$ and $\kv_2$ with $0\le |\kv_{1(2)}| \le k_F$, where
$k_F$ is the Fermi wavevector.
The probability $p(k)$ is then\cite{Ov_tesi,3p}
\begin{equation}
p(k)=24\frac{k^2}{k_F^3}-36\frac{k^3}{k_F^4}+12\frac{k^5}{k_F^6},
\label{eq_pk}
\end{equation}
with $k$ ranging from $0$ to $k_F$. ($p(k)$ vanishes at
the endpoints of its domain, maximizes around the middle, and integrates to
1.) Then we find analytically that Eq.~(\ref{eq_guunonint})
gives the correct result for the non-interacting uniform gas.
Numerically, in the range $0\le r/r_s\le 3$ and with 
a truncation of the infinite sum over $\ell$ at $\ell_{\rm max}=7$,
Eq.~(\ref{eq_guunonint}) reproduces the known 
exchange-only $g_{\uu}$ within an accuracy of $10^{-6}$.

With these simple concepts in mind, we can now proceed to compute
an interacting pair-distribution function by introducing a suitable
electron-electron potential which describes the interactions
in a uniform electron gas. Then we just have to replace
the spherical Bessel functions $j_{\ell}$ in Eqs.~(\ref{eq_gudnonint}) 
and~(\ref{eq_guunonint}) with the functions $R_{\ell}$, solutions
of the radial Schr\"odinger equation with the chosen two-body potential.
Unless this potential is very sophisticated, such a treatment 
will fail to describe long-range correlations, which are mainly
governed by collective modes, and will fail to satisy
the particle-conservation sum rule on $g_{\sigma\sigma'}(r)$ (Eq. (47) of 
Ref.~\cite{PW}).

\section{Solution using the Overhauser screened Coulomb potential}
\label{sec_ovpot}
Overhauser~\cite{Ov} proposed a simple and reasonable model
for the screened Coulomb repulsion $V(r)$ in the uniform electron gas.
In the standard uniform-electron-gas
model, a rigid positively-charged background maintains
electrical neutrality. Thus Overhauser took the
sphere of volume $n^{-1}$ as the boundary within which the screening
charge density is $ne$ and outside of which it is zero. This is equivalent
to assuming that the probability of finding three electrons in a sphere
of radius $r_s$ is exactly zero, an assumption which is nearly
true. In fact, numerical estimates of this probability
for an interacting electron gas
show that it is indeed small.\cite{3p} (At $r_s=5$ 
the ratio between the probabilities of finding three
and two electrons in the same sphere of radius $r_s$ is about $1/11$; 
for larger $r_s$ this ratio is lower, and for smaller $r_s$ it is higher, being
about $1/7$ at $r_s=0$.) 

Thus, for interelectronic distances $r<r_s$ we expect
 the Overhauser potential to be close to the true potential felt
by an electron moving in a uniform electron gas when another
electron is fixed at the origin. In the region $r>r_s$ the potential is set to
zero, and so is not expected to be reliable. 

We also expect to have results that become more accurate as the density
decreases, since the probability of having three electrons in the
same sphere of radius $r_s$ becomes lower and lower.
Finally, at high and intermediate densities our results will be much closer
to the true $g(r)$ for antiparallel-spin correlations than for 
parallel-spin ones. When two electrons of opposite spins 
are in the same sphere of
radius $r_s$, a third electron is excluded from the
sphere because of both the Pauli principle and the Coulomb repulsion.
For a pair of parallel-spin electrons, only the Coulomb repulsion prevents
a third electron of opposite spin
from entering the sphere of radius $r_s$, a mechanism
which becomes less efficient as the density 
(and thus the kinetic energy) increases.

\begin{figure}
\includegraphics[width=\columnwidth]{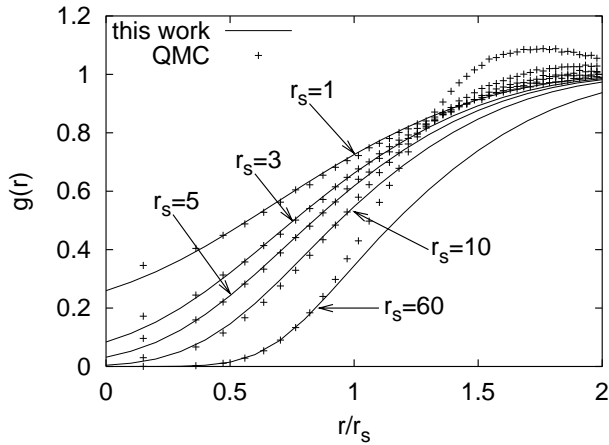} 
\caption{Pair-distribution functions $g(r)$ for the uniform
electron gas computed with the Overhauser~\cite{Ov} potential
for different values of the electron-density parameter $r_s$. Also
shown are the Quantum Monte Carlo data from Ref.~\cite{OHB}.}
\label{fig_totg}
\end{figure}

In atomic units, the Overhauser potential is given by
\begin{eqnarray}
  V(r)    = &  \frac{1}{r_s}\left(\frac{r_s}{r}+\frac{r^2}{2r_s^2}-\frac{3}{2}
\right)  \qquad  &  (r\le r_s) \nonumber  \\
  V(r)    =  & 0    & (r>r_s). \label{eq_potOv}
\end{eqnarray}  
Defining $s=r/r_s$, $q=k\,r_s$ and $u_{\ell}(s)=qsR_{\ell}(s)$, 
the corresponding radial Schr\"odinger equation reads
\begin{eqnarray}
& & \left[\tfrac{d^2}{ds^2}-\tfrac{\ell(\ell+1)}{s^2}+q^2
-r_s\left(\tfrac{1}{s}+\tfrac{s^2}{2}-\tfrac{3}{2}\right)\right]
u_{\ell}=0 \qquad (s\le 1) \nonumber \\
& & \left[\tfrac{d^2}{ds^2}-\tfrac{\ell(\ell+1)}{s^2}+q^2
\right]u_{\ell} = 0 \qquad (s>1). 
\label{eq_schro}
\end{eqnarray}
In order to find $u_{\ell}(s)$, we proceed as follows. 
For $0\le s < 1$, we expand $u_{\ell}$ in a power series:
\begin{equation}
u_{\ell}(s)=\alpha_{\ell}\sum_{n=\ell+1}^{\infty}c_n^{(\ell)}\,
s^n \qquad (0\le s <1).
\label{eq_u1}
\end{equation}
Inserting this expansion into the corresponding radial equation, we find
a recursion relation between the coefficients $c_n^{(\ell)}$,
\begin{eqnarray}
& & c_n^{(\ell)} = \frac{\tfrac{r_s}{2}
c_{n-4}^{(\ell)}-\left(\tfrac{3}{2}r_s+q^2\right)c_{n-2}^{(\ell)}
+r_sc_{n-1}^{(\ell)}}{(n+\ell)(n-\ell-1)}  
\;\;\;\; (n>\ell+1) \nonumber \\
& & c_{\ell+1}^{(\ell)}=1, \qquad \qquad  c_{n}^{(\ell)}=0
\;\;\; (n\le \ell).
\end{eqnarray}
For $s> 1$ the solution is $qs$ times the spherical Bessel function $j_{\ell}$
with a phase shift $z_{\ell}$. It can be usefully written as
\begin{eqnarray}
u_{\ell}(s) &  = &  qs\big[f_{\ell}(qs)\sin(qs-z_{\ell})+
(-1)^{\ell+1}f_{-\ell-1}(qs)\times \nonumber \\
& & \cos(qs-z_{\ell})\big]
 \qquad \qquad \qquad (s >1),
\label{eq_u2}
\end{eqnarray}
where $f_j(x)$ are given by
\begin{eqnarray}
f_0(x)=x^{-1} \qquad f_1(x)=x^{-2} \nonumber \\
f_{j-1}(x)+f_{j+1}(x)=(2j+1)\,x^{-1}f_j(x).
\label{eq_fl}
\end{eqnarray}
Equation~(\ref{eq_u2}) is properly normalized, since
the noninteracting case of Eqs.~(\ref{eq_gudnonint}) and~(\ref{eq_guunonint})
is exactly recovered when $z_{\ell}=0$.
By matching $u_{\ell}(s)$ and its first derivative at $s=1$, we can
find $\alpha_{\ell}$ and the phase shift $z_{\ell}$. They are given
by
\begin{eqnarray}
& & \frac{\alpha_{\ell}^2}{q^2}  = 
 \frac{q^2\left[f_{\ell}(q)f_{-\ell}(q)
+f_{\ell-1}(q)f_{-\ell-1}(q)\right]^2}{\mathcal{S}_{\ell}(q,r_s)^2+
\mathcal{C}_{\ell}(q,r_s)^2} \label{eq_alpha} \\
& &  \tan(q-z_{\ell}) =  (-1)^{\ell}\frac{\mathcal{S}_{\ell}(q,r_s)}
{\mathcal{C}_{\ell}(q,r_s)} \label{eq_z},
\label{eq_tgz}
\end{eqnarray}
where 
\begin{eqnarray}
\mathcal{S}_{\ell}=qf_{-\ell}(q)B_{\ell}+
f_{-\ell-1}(q)A_{\ell} \\
\mathcal{C}_{\ell}=
f_{\ell}(q)A_{\ell}-qf_{\ell-1}(q)B_{\ell} \\
A_{\ell}=A_{\ell}(q,r_s)=\sum_{n=\ell+1}^{\infty}c_n^{(\ell)}(n+\ell) 
\label{eq_A}\\
B_{\ell}=B_{\ell}(q,r_s)=\sum_{n=\ell+1}^{\infty}c_n^{(\ell)}.
\label{eq_B}
\end{eqnarray}
With 8 ($\ell_{\rm max}=7$) partial waves and 
the infinite sum over $n$ 
in Eqs.~(\ref{eq_u1}), (\ref{eq_A}) and~(\ref{eq_B})
truncated at $n_{\rm max}\sim 20$,
we reach accurate convergence for $0\le r/r_s\le 2$. 

We thus obtain smooth radial functions $R_{\ell}(s,r_s,q)$ to
replace the functions $j_{\ell}$ in Eqs.~(\ref{eq_gudnonint})
and~(\ref{eq_guunonint}). Then we have to average over $q$
to obtain $g_{\sigma\sigma'}(s,r_s)$. We used the 
probability distribution $p(q)$ corresponding to an unpolarized
noninteracting
uniform electron gas, i.e., we took Eq.~(\ref{eq_pk}) with $k_F=
\frac{1}{r_s}(\frac{9\pi}{4})^{1/3}$,
\begin{equation}
p(q)=\tfrac{16}{3\pi}q^2\left[2-\left(\tfrac{12}{\pi}\right)^{1/3}q
+\tfrac{4}{9\pi}\,q^3\right],
\label{eq_pq}
\end{equation}
with $q$ ranging from $0$ to $(9\pi/4)^{1/3}$. The $q$-dependence
of our results is rather weak, so we expect to have no significant
change if we use an interacting momentum distribution instead of
the one of Eq.~(\ref{eq_pq}).

The results for
the total $g(r)$ are shown in Fig.~\ref{fig_totg}, together
with the newest QMC results from Ref.~\cite{OHB}.
We find $g(r)$ in accurate agreement with the QMC
data for $0.5\lesssim r/r_s\lesssim 1$ for a wide 
range of electron densities. In the shortest-range region,
 $r/r_s \lesssim 0.5$, the QMC data
are known to suffer large errors and are not so reliable.
(In this region there is in fact a significant discrepancy between the
data from Ref.~\cite{OHB} and those from Ref.~\cite{cepald}.)
We believe that for $r/r_s \lesssim 0.5$ the present treatment 
provides results much closer to the true $g(r)$. As said, for $r/r_s>1$
the results obtained with the Overhauser potential are not reliable.

\section{Short-range coefficients of the pair-distribution function}
The small-$s$ expansion of the spin-resolved $g_{\ud}$ and
$g_{\uu}$ (where $s=r/r_s$),
\begin{eqnarray}
& & g_{\ud}(s,r_s)  =  a_0^{\ud}(r_s)+a_1^{\ud}(r_s)s+a_2^{\ud}(r_s)s^2+O(s^3)
\label{eq_expgud}\\
& &  g_{\uu}(s,r_s)  =  a_2^{\uu}(r_s) s^2 + a_3^{\uu}(r_s) s^3+ O(s^4),
\label{eq_expguu}
\end{eqnarray}
obtained by solving the Overhauser equation, has coefficients
\begin{eqnarray}
a_0^{\ud} & = &  \left\langle\frac{\alpha_0^2}{q^2}\right\rangle 
\label{eq_a0ud}\\
a_1^{\ud} & = & r_s\left\langle\frac{\alpha_0^2}{q^2}\right\rangle =
r_s a_0^{\ud} \label{eq_a1ud} \\
a_2^{\ud} & = & \frac{r_s}{12}(5 r_s-6)\left\langle
\frac{\alpha_0^2}{q^2}\right\rangle-\frac{\langle\alpha_0^2\rangle}{3}
+3\left\langle\frac{\alpha_1^2}{q^2}\right\rangle \label{eq_a2ud}\\
a_2^{\uu} & = & 6\left\langle\frac{\alpha_1^2}{q^2}\right\rangle 
\label{eq_a2uu}\\
a_3^{\uu} & = & 3\,r_s\left\langle\frac{\alpha_1^2}{q^2}\right\rangle=
\frac{r_s}{2}a_2^{\uu},
\label{eq_a3uu}
\end{eqnarray}
where the $\alpha_{\ell}$ are given by Eq.~(\ref{eq_alpha}) and
$\langle\rangle$ means average over $p(q)$ of Eq.~(\ref{eq_pq}).
Equations~(\ref{eq_a1ud}) and~(\ref{eq_a3uu}) are
the known cusp conditions,\cite{Kimball2,Kimball}
due to the dominance for $r\to 0$ of the term $1/r$ in the
Schr\"odinger equation. 
\begin{figure}
\includegraphics[width=\columnwidth]{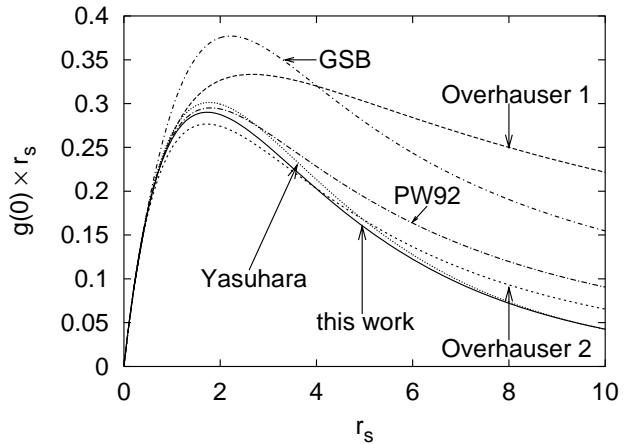} 
\caption{Pair-distribution function at zero-interelectronic
distance multiplied by the density parameter $r_s$. 
The present result (exact solution of the Overhauser model)
is compared with the two Overhauser
formulas\cite{Ov} (1 and 2 iterations), with the 
Yasuhara\cite{Yasuhara} electron-electron ladder evaluation, with
the Perdew and Wang\cite{PW} (PW92) formula, and with the extrapolation from
QMC data\cite{OHB} obtained by Gori-Giorgi, Sacchetti and 
Bachelet\cite{GSB} (GSB).}
\label{fig_g0}
\end{figure}

Equation~(\ref{eq_a0ud}) gives the value
of the pair-distribution function at zero-interelectronic distance,
$g(0)=a_0^{\ud}/2$. In Fig.~\ref{fig_g0} we compare the present
result with other estimates of $g(0)$: (i) two approximate solutions
that Overhauser\cite{Ov} obtained from his model potential
after setting $q=0$ in the
radial Schr\"odinger equation for $s\le 1$; (ii) the Yasuhara\cite{Yasuhara} 
electron-electron ladder evaluation; (iii) the Perdew-Wang\cite{PW} (PW92)
formula, and (iv) the result extrapolated from QMC data\cite{OHB} by
Gori-Giorgi, Sacchetti and Bachelet\cite{GSB} (GSB). 
The value of $g(0)$ has been multiplied by the density parameter
$r_s$ in order to amplify the differences between the curves, which are
mainly located at low densities. We see that
the first Overhauser formula, obtained from a first iteration,
is quite far from the ``exact'' solution of his model presented in
this work. His second formula, derived by a second iteration, is much
closer to our solution (within $5\%$ at $r_s=2$, $50\%$ at $r_s=10$).
The Yasuhara electron-electron ladder evaluation is the closest to our result
for $r_s\gtrsim 5$ (e.g., the difference is only $0.6\%$ at $r_s=10$). This is
not surprising since both calculations are more reliable at lower densities.
The PW92 formula is mainly a
Pad\'e approximation of the Yasuhara values; it is not very
accurate at densities $r_s\gtrsim 4$, where the electron-electron ladder
$g(0)$ goes to zero as\cite{BPE} $r_s^{3/2}{\rm e}^{-A\sqrt{r_s}}$, while
the PW92 Pad\'e approximant goes as $r_s^{-3}$. However, the 
PW92 formula reproduces quite accurately the Yasuhara values at
metallic densities, and has the advantage of being very simple and
handy for many purposes.
The extrapolated GSB value is, for $r_s\gtrsim 0.7$, 
much higher than the curve obtained in the present
work, as expected from the results of Fig.~\ref{fig_totg}, in which our
$g(r)$ curves always lie below the QMC data for $r/r_s\lesssim 0.3$.

The high-density limit of our $a_0^{\ud}(r_s)$ is
\begin{equation}
a_0^{\ud}(r_s\to 0)= 1-0.684\,r_s+O(r_s^2),
\label{eq_a0udHD}
\end{equation}
which differs from the exact one,\cite{Geldart,Kg0}
$1-0.7317\,r_s$, by only 6.5\% in the coefficient of $r_s$
and is slightly better than the high-density limit of the 
Yasuhara\cite{Yasuhara} $a_0^{\ud}$, $1 - 0.663r_s$.\cite{BPE}
We can conclude that the value of $g(0)$ obtained by solving the Overhauser 
model is very reliable: it almost recovers the 
exact high-density limit, and agrees in the
low-density limit with the electron-electron ladder evaluation
of Yasuhara.
In the latter limit, where the electrons become strictly correlated,\cite{3p}
$a_2^{\ud}$ and $a_2^{\uu}$ tend to zero.
\begin{figure}
\includegraphics[width=\columnwidth]{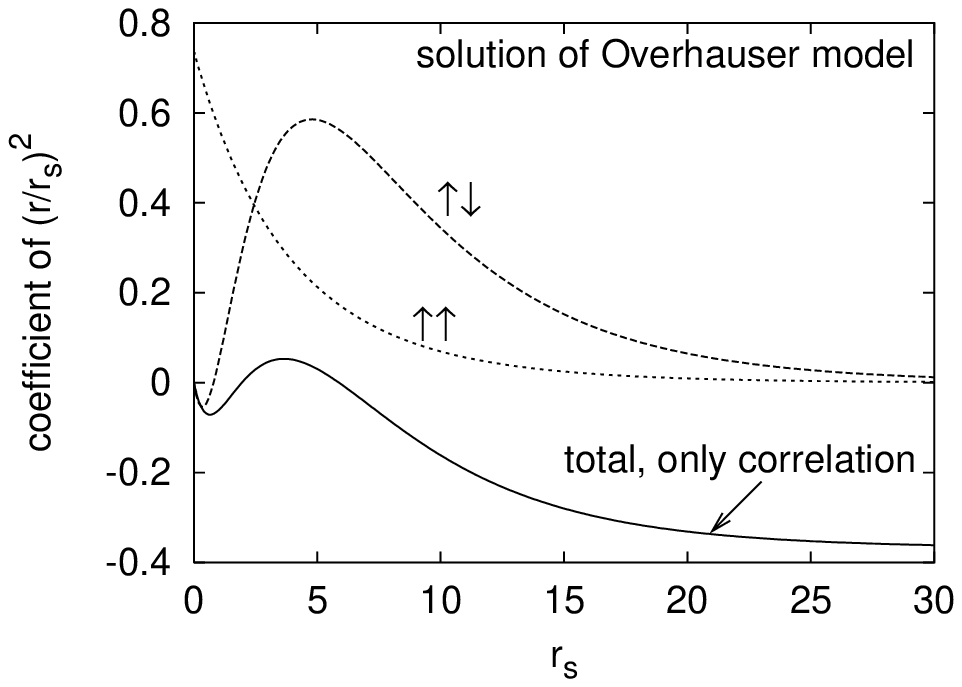} 
\includegraphics[width=\columnwidth]{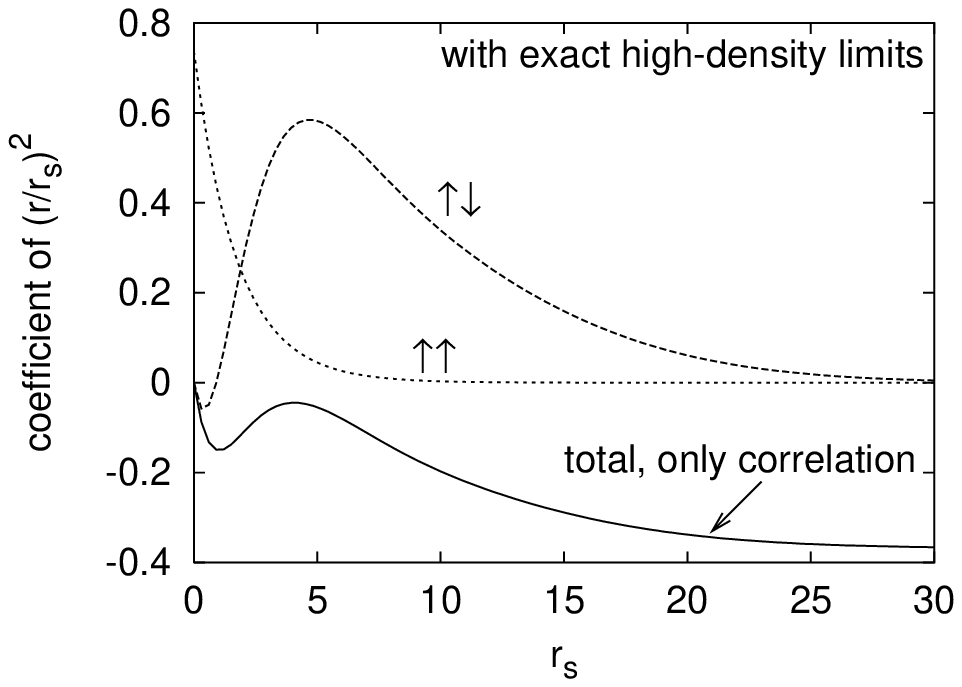} 
\caption{Coefficients of $(r/r_s)^2$ in the
small-$r$ expansion of the pair-distribution function. The
parallel- and antiparallel-spin contributions are separately
shown. The total coefficient for the correlation part, 
i.e., $a_2^c=
\tfrac{1}{2}(a_2^{\ud}+a_2^{\uu})-\tfrac{1}{10}(\tfrac{9\pi}{4})^{2/3}$,
is also reported. Upper panel: solution of the Overhauser model; lower
panel: interpolation formulas including the exact high-density limits.}
\label{fig_a2}
\end{figure}

Our results for $g(0)$ can be usefully parametrized with the following form
\begin{equation}
a_0^{\ud}(r_s)=(1-Br_s+Cr_s^2+Dr_s^3+Er_s^4)\,{\rm e}^{-dr_s}.
\label{eq_fitg0}
\end{equation} 
We constrain our fit to reproduce the exact high-density
limit, in order to be as close as possible to the true $g(0)$. Thus we
fix $B=0.7317-d$. A best fit of the remaining 4 free parameters to
our result for $r_s\le 10$ gives $C=0.08193$, $D=-0.01277$,
$E=0.001859$, $d=0.7524$. Then $B=-0.0207$. On the scale of Fig.~\ref{fig_g0},
the fit error is invisible.

As far as we know,
the coefficients $a_2^{\ud}$ and $a_2^{\uu}$, given
by Eqs.~(\ref{eq_a2ud}) and~(\ref{eq_a2uu}),
are presented in this work for the first time. The good performance of
$g(0)$ obtained with the Overhauser potential, together with the
agreement of $g(r)$ with the QMC data for $0.5\lesssim r/r_s \lesssim 1$, 
suggest that our calculation of these coefficients is quantitatively 
reliable. In the upper panel of
Fig.~\ref{fig_a2}, we report $a_2^{\ud}$ and $a_2^{\uu}$ as
a function of $r_s$. While the parallel-spin part shows a predictable
dependence on $r_s$, decreasing monotonically from the noninteracting
value $\frac{1}{5}(\frac{9\pi}{4})^{2/3}$ to zero as $r_s$ increases,
the $\ud$ part presents at first glance a non-intuitive dependence on the
density. However, if we consider the known 
high- and low-density behavior of $a_2^{\ud}$, we see that we should
expect a function which is qualitatively similar to the one
we obtained. In fact, in the $r_s\to 0$ limit $a_2^{\ud}$ is known
to approach zero from negative values,\cite{reply} while in the
$r_s\to\infty$ limit we expect $a_2^{\ud}$ to approach zero asymptotically
from positive values, in order to fulfill the positivity constraint on
$g_{\ud}(r)$. Thus, $a_2^{\ud}$ must have a minimum and a maximum, and
must cross zero in between, as does our result. 

The high-density limit of our $s^2$ coefficients is
\begin{eqnarray}
a_2^{\ud}(r_s\to 0) & = & -0.34\,r_s +O(r_s^2)\label{eq_a2udHD} \\
a_2^{\uu}(r_s\to 0) & = & \tfrac{1}{5}(\tfrac{9\pi}{4})^{2/3}
-0.192\,r_s+O(r_s^2). \label{eq_a2uuHD}
\end{eqnarray}
The $\ud$ value is in excellent agreement with the exact
one,\cite{reply} $-0.3356\,r_s$, while the $\uu$ coefficient of
$r_s$ in the r.h.s. of Eq.~(\ref{eq_a2uuHD}) is 54\% 
smaller than the exact one,\cite{reply} $-0.422$; see
also Ref.~\cite{Kimball2}.
As said, in the high-density limit the assumption 
of zero probability of having a third electron in the same sphere 
of radius $r_s$ can be nearly true only for antiparallel spin
electrons, since in this case the Pauli principle keeps the third
electron away. So, our $a_2^{\ud}$ should be very close to the true
one, while the $\uparrow\uparrow$ short-range correlations are
underestimated at higher densities in the present treatment
(i.e., the true $a_2^{\uu}$ deviates more
from the noninteracting value $\frac{1}{5}(\frac{9\pi}{4})^{2/3}$, so it
is lower than our result). 
However, $a_2^{\uu}$ has
a very simple dependence on $r_s$, so, as we shall see,
 it is very easy to write down an interpolation formula between the high-
and the low-density limits. 

In the upper panel of Fig.~\ref{fig_a2} we also show $a_2^c$, the total 
$(r/r_s)^2$ coefficient of the correlation part of $g(r)$,
i.e., $\frac{1}{2}(a_2^{\ud}+a_2^{\uu})$ minus the noninteracting
value $\frac{1}{10}(\frac{9\pi}{4})^{2/3}$. We find that $a_2^c$
is positive in the density range $1.94<r_s<5.88$, 
and negative elsewhere. However, we believe that the positive region
is a consequence of underestimating the $\uu$ correlations. 


Our results are very well reproduced by the simple formulas
\begin{eqnarray}
a_2^{\ud} & = & (-\beta_a r_s+\gamma_a r_s^2+\delta_a r_s^3
+\epsilon_a r_s^4)\,{\mathrm e}^{-d_a r_s} \label{eq_a2udfit} \\
a_2^{\uu} & = & \tfrac{1}{5}\left(\tfrac{9\pi}{4}\right)^{2/3}
(1-\beta_p r_s+\gamma_p r_s^2)\,{\mathrm e}^{-d_p r_s},
\label{eq_a2uufit}
\end{eqnarray}
where $\beta_a=0.32$, $\gamma_a=0.4069$, $\delta_a=-0.04455$,
$\epsilon_a=0.003064$, $d_a=0.4235$, $\beta_p=0.01624$,
$\gamma_p=0.00264$, $d_p=0.2456$.

However, as said, the $\uu$ correlations are underestimated. We can
use the formulas of Eqs.~(\ref{eq_a2udfit}) and~(\ref{eq_a2uufit})
to recover the exact high-density limits,\cite{reply} by simply changing
the parameters $\beta_a$ and $d_p$. The new values are $\beta_a=0.3356$,
$d_p=0.5566$. In this way, the $\ud$ coefficient is basically unchanged,
while the $\uu$ is lowered because of the exact slope 
at $r_s=0$, and should be much closer
to the true one. The new $a_2$ coefficients are shown in the lower panel of
Fig.~\ref{fig_a2}, together with the total coefficient for 
the correlation part. We see that now $a_2^c$ is always negative.

Ziesche\cite{Z2000} has proposed $1 - 5(\frac{9\pi}{4})^{-2/3}a_2^{\uu}$
as a measure of correlation strength, varying monotonically
from 0 for $r_s\to 0$ to 1 for $r_s\to \infty$.

\section{Coupling-constant average}
The coupling-constant average\cite{OT} of the pair-distribution
function for the uniform electron gas is given by
\begin{equation}
\overline{g}(s,r_s)=\tfrac{1}{r_s}\int_0^{r_s}g(s,r_s')\,dr_s',
\label{eq_defgbar}
\end{equation}
where $s$ is kept fixed.
The function $\overline{g}$ is important because it can account
for the kinetic energy of correlation. Its short-range coefficients
can be computed by integrating the simple formulas
of Eqs.~(\ref{eq_fitg0}), (\ref{eq_a2udfit}), and~(\ref{eq_a2uufit}):
\begin{eqnarray}
\overline{a}_0^{\ud} & = & \tfrac{1}{r_s}[(-\overline{A}+
\overline{B}r_s+\overline{C}r_s^2+\overline{D}r_s^3+\overline{E}r_s^4)\,
{\rm e}^{-dr_s} \nonumber \\ 
& & +\overline{A}] \label{eq_a0av} \\
\overline{a}_1^{\ud} & = & \tfrac{1}{r_s}[(-\overline{A}_1+
\overline{B}_1r_s+\overline{C}_1r_s^2+\overline{D}_1r_s^3
+\overline{E}_1r_s^4 \nonumber \\ 
& & +\overline{F}_1r_s^5)\,{\rm e}^{-dr_s}+\overline{A}_1]
\label{eq_a1av} \\
\overline{a}_2^{\ud} & = & \tfrac{1}{r_s}[(-A_a+B_a r_s
+C_a r_s^2+D_a r_s^3+E_a r_s^4)\,{\mathrm e}^{-d_a r_s} \nonumber \\
& & +A_a]
\label{eq_audav} \\
\overline{a}_2^{\uu} & = & \tfrac{1}{5}\left(\tfrac{9\pi}{4}\right)^{2/3}
\tfrac{1}{r_s}[(-A_p+B_p r_s
+C_p r_s^2)\,{\mathrm e}^{-d_p r_s} \nonumber \\
& & +A_p],
\label{eq_auuav}
\end{eqnarray}
where for $\overline{a}_0^{\ud}$ we obtain
$\overline{A}=1.696$, $\overline{B}=-0.2763$, $\overline{C}=-0.09359$,
$\overline{D}=0.003837$, $\overline{E}=-0.002471$, and 
for $\overline{a}_1^{\ud}$
we obtain $\overline{A}_1=3.356$, $\overline{B}_1=-2.525$, $\overline{C}_1=
-0.45$, $\overline{D}_1=-0.106$, $\overline{E}_1=0.000553$,
$\overline{F}_1=-0.00247$. Equation~(\ref{eq_a1av}) has been obtained
by integrating $r_s a_0^{\ud}(r_s)$, according to the cusp condition. In this
way, we are able to give short-range coefficients which exactly satisfy 
the modified cusp condition for $\overline{g}(r)$. (In the 
Perdew-Wang\cite{PW} model for $\overline{g}$, the cusp condition is
accurately but not exactly satisfied.)

For the $s^2$ coefficients, using the values corresponding to the exact
high-density limit formulas, we have:
$A_a=5.9313$, $B_a=-2.5119$, $C_a=-0.6997$, $D_a=0.03686$,
$E_a=-0.007235$, $d_a=0.4235$, $A_p=1.775$, $B_p=0.0121$, $C_p=-0.00474$,
$d_p=0.5566$.

\section{Spin-polarized gas}
The spin-polarized electron gas is characterized by the parameter
$\zeta=(n_{\uparrow}-n_{\downarrow})/n$, where $n_{\uparrow}$
and $n_{\downarrow}$ are the densities of spin-up and -down electrons.
The pair-distribution function averaged over  $\uu$, $\dd$, and $\ud$ 
pairs is then
\begin{eqnarray}
g(r,r_s,\zeta) & = &\left(\tfrac{1+\zeta}{2}\right)^2g_{\uu}(r,r_s,\zeta)+
\left(\tfrac{1-\zeta}{2}\right)^2g_{\dd}(r,r_s,\zeta) \nonumber \\
& & +\tfrac{(1-\zeta^2)}{2}g_{\ud}(r,r_s,\zeta).
\end{eqnarray}
Here we shall motivate and use an approximate spin scaling relation for the
short-range part of $g$:
\begin{equation}
g_{\sigma\sigma'}(r,r_s,\zeta)=g_{\sigma\sigma'}(r,r_s^{\sigma\sigma'}(\zeta),
\zeta=0),
\end{equation}
where $r_s^{\uu}=r_s\,(1+\zeta)^{-1/3}$, $r_s^{\dd}=r_s\,(1-\zeta)^{-1/3}$, and
$r_s^{\ud}=2\,r_s/[(1+\zeta)^{1/3}+(1-\zeta)^{1/3}]$.

In the present treatment, information about the spin polarization
of the system enters when we average over the probability $p(q)$.
For a spin-unpolarized ($\zeta=0$) gas, the correct $p(q)$ is the one 
of Eq.~(\ref{eq_pq}). For a fully polarized ($\zeta=1$) gas, we can
similarly
obtain $p(q)$ from Eq.~(\ref{eq_pk}) by using the proper relation between
$k_F$ and $r_s$, i.e., $k_F^{\zeta=1}=\frac{1}{r_s}(\frac{9\pi}{2})
^{1/3}=2^{1/3}k_F^{\zeta=0}$. For a partially-polarized gas 
($0<\zeta<1$), we have to distinguish
between the parallel- and the antiparallel-spin cases. In the
parallel-spin case, we
can still obtain $p(q)$ from Eq.~(\ref{eq_pk}) by using $k_{F\uparrow}=
(1+\zeta)^{1/3}k_F^{\zeta=0}$ for $\uu$ interactions, and $k_{F\downarrow}=
(1-\zeta)^{1/3}k_F^{\zeta=0}$ for $\dd$ interactions. This means that
we can obtain $p_{\zeta}^{\uu}(q)$ and $p_{\zeta}^{\dd}(q)$ by
rescaling $p(q)$ of Eq.~(\ref{eq_pq}) as follows:
\begin{eqnarray}
p_{\zeta}^{\uu}(q)=(1+\zeta)^{-1/3}\,p\left[q\,(1+\zeta)^{-1/3}\right] 
\label{eq_pquu}\\
p_{\zeta}^{\dd}(q)=(1-\zeta)^{-1/3}\,p\left[q\,(1-\zeta)^{-1/3}\right],
\label{eq_pqdd}
\end{eqnarray} 
with $q$ ranging from $0$ to $[\frac{9\pi}{4}(1+\zeta)]^{1/3}$ for
the $\uu$ case and from $0$ to $[\frac{9\pi}{4}(1-\zeta)]^{1/3}$
for the $\dd$ case. 
In the case of a pair of antiparallel-spin electrons, 
we have to compute geometrically 
the probability distribution for a variable $k=\frac{1}{2}|\kv_2-\kv_1|$
when $0\le|\kv_1|\le k_{F\downarrow}$ and $0\le|\kv_2|\le k_{F\uparrow}$. 
This gives
\begin{eqnarray}
p_{\zeta}^{\ud}(k) & & = 24\tfrac{k^2}{k_{F\uparrow}^3}  
\qquad ({\rm for}\; 0\le k\le k_-) \nonumber \\
 & & =-\tfrac{9}{4}\tfrac{(k_{F\uparrow}^2-k_{F\downarrow}^2)^2}{
k_{F\uparrow}^3k_{F\downarrow}^3}k+12\tfrac{(k_{F\uparrow}^3+
k_{F\downarrow}^3)}{k_{F\uparrow}^3k_{F\downarrow}^3}k^2-
18\tfrac{(k_{F\uparrow}^2+k_{F\downarrow}^2)}{k_{F\uparrow}^3k_{F\downarrow}
^3}k^3 \nonumber \\
& & +\tfrac{12}{k_{F\uparrow}^3k_{F\downarrow}^3}k^5
 \qquad ({\rm for}\; k_- \le
k \le k_+),
\end{eqnarray}
where $k_{\pm}=\frac{1}{2}(k_{F\uparrow}\pm k_{F\downarrow})$.
Our goal is to write down a scaling relation which allows us to
use the results obtained for the $\zeta=0$ gas. To do this, first
we have to write down an approximate $p_{\zeta}^{\ud}(q)$
obtained by scaling $p(q)$. A reasonable choice, having correct limits 
and normalization, is
\begin{equation}
p_{\zeta}^{\ud}(q)\approx \tfrac{2}{(1+\zeta)^{1/3}+(1-\zeta)^{1/3}}
\,p\left[\tfrac{2\,q}{(1+\zeta)^{1/3}+(1-\zeta)^{1/3}}\right],
\label{eq_pqud}
\end{equation}
with $q$ varying
from $0$ to $(\frac{9\pi}{4})^{1/3}\frac{[(1+\zeta)^{1/3}+
(1-\zeta)^{1/3}]}{2}$.

In order to use the results obtained for the unpolarized gas, we now
consider that $q=kr_s$. Scaling $q$ according to Eqs.~(\ref{eq_pquu}),
(\ref{eq_pqdd}) and~(\ref{eq_pqud}) can thus be approximately
equivalent to rescaling $r_s$. This is exact if the
Overhauser radial Schr\"odinger equation is changed in the same way when
we replace $q$ with $\mu q$ and when we replace $r_s$ with $r_s/\mu$. In the
latter case, one also has to consider that the variable $s=r/r_s$ is
changed into $s'=\mu r/r_s=\mu s$. One can easily check that this 
scaling property is satisfied by the equation with the potential
set to zero ($s>1$), which means that it is exact at the exchange-only level.
For the case of nonzero potential  ($0\le s\le 1$),
the scaling condition is satisfied only 
by the kinetic terms and by the 
$1/s$ part of the potential. However, since for small $s$ the $1/s$ part
of the potential is dominant, we expect to obtain rather good results
by applying this scaling to the short-range coefficients.

When we apply our scaling relation to the total $s^2$ coefficient of $g$, 
we obtain
\begin{eqnarray}
a_2(r_s,\zeta)& & = \left(\tfrac{1+\zeta}{2}\right)^2(1+\zeta)^{2/3}a_2^{\uu}
[r_s(1+\zeta)^{-1/3}]+ \nonumber \\
& &  \left(\tfrac{1-\zeta}{2}\right)^2(1-\zeta)^{2/3}a_2^{\uu}
[r_s(1-\zeta)^{-1/3}]+\tfrac{(1-\zeta^2)}{2}\times \nonumber \\
& & \left[\tfrac{(1+\zeta)^{1/3}+(1
-\zeta)^{1/3}}{2}\right]^2a_2^{\ud}
\left[\tfrac{2r_s}{(1+\zeta)^{1/3}+(1-\zeta)^{1/3}}\right],
\label{eq_a2zeta}
\end{eqnarray}   
where the functions $a_2^{\ud}(r_s)$ and $a_2^{\uu}(r_s)$
can be either the ones of Eqs.~(\ref{eq_a2udfit}) and~(\ref{eq_a2uufit})
or the ones of Eqs.~(\ref{eq_audav}) and~(\ref{eq_auuav}), depending
whether one is interested in $g$ or in its coupling-constant average
$\overline{g}$. Equation~(\ref{eq_a2zeta}) is
exact (beyond the Overhauser model) in several limits: for exchange, for 
parallel-spin correlation in the high-density limit (Eq. (10) of
Ref.\cite{reply}), and for the low-density limit. 
The fact that Eq.~(\ref{eq_a2zeta}) is exact for parallel-spin correlations
in the $r_s\to 0$ limit suggests that
the results for the short-range coefficients of $g(r)$ are not
affected by
the violation of the scaling
relation by the part of the potential which remains finite when $s\to 0$.

\begin{figure}
\includegraphics[width=\columnwidth]{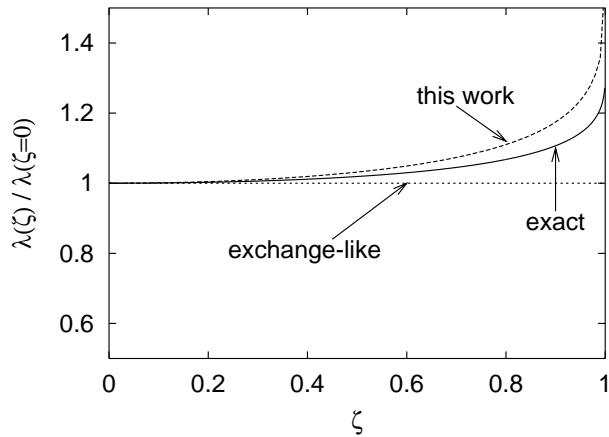} 
\caption{Spin-polarization dependence of $g(0)$ at high-
density. $\lambda(\zeta)$ is defined by Eq.~(\ref{eq_exl}).
The scaling prediction of Eq.~(\ref{eq_lambdaz}) is compared to the exact
$\zeta$-dependence,\cite{Rassolov,BPE} and to the exchange-like
scaling adopted in Refs.~\cite{PW} and~\cite{MQ}.}
\label{fig_g0zHD}
\end{figure}

For the $\ud$ interactions, we have also the error due to the
use of an approximate relative-momentum distribution.
We can get an idea of this error by applying our 
spin-scaling relation to $g(0)$, with the result
\begin{equation}
g(r=0,r_s,\zeta)=(1-\zeta^2)g(r=0,
\tfrac{2r_s}{(1+\zeta)^{1/3}+(1-\zeta)^{1/3}},\zeta=0).
\label{eq_g0z}
\end{equation}
Equation~(\ref{eq_g0z}) is different from that of
Marinescu and Quinn\cite{MQ} or Perdew and Wang,\cite{PW}
obtained by using an exchange-like ($r_s=0$)
$\zeta$-dependence. The high-density
limit of $g(0,r_s,\zeta)$ is
\begin{equation}
g(r=0,r_s\to 0,\zeta)=\tfrac{1}{2}(1-\zeta^2)\,[1-\lambda(\zeta)\,r_s],
\label{eq_exl}
\end{equation}
where from Eq.~(\ref{eq_g0z})
\begin{equation}
\lambda(\zeta)=\frac{2\,\lambda(\zeta=0)}{(1+\zeta)^{1/3}+
(1-\zeta)^{1/3}}.
\label{eq_lambdaz}
\end{equation}
In Fig.~\ref{fig_g0zHD} we compare our Eq.~(\ref{eq_lambdaz})
with the exact $\lambda(\zeta)/\lambda(0)$,\cite{Rassolov,BPE} 
and with the exchange-like
approximation of Refs.~\cite{PW} and~\cite{MQ}, in which this ratio is $1$.
We see that our scaling relation is always closer to the exact
$\zeta$-dependence than is
the exchange-like one. The maximum error is at $\zeta=1$,
where our ratio is $1.587$ and the exact one is $1.332$.
(By the way, we numerically confirm the
equivalence of the exact expressions in Refs.~\cite{Rassolov} 
and~\cite{BPE} for all $\zeta$.)  

\section{Conclusions and future directions}
Our work supports and extends Overhauser's picture of short-range
correlations in the uniform electron gas of density $n=3/4\pi r_s^3$.
When two electrons are separated by a distance $r\lesssim r_s$, there is
very little probability of finding a third electron within a sphere of
radius $r_s$ about the first (especially in the strong-interaction or
large-$r_s$ limit). Thus the short-range correlations arise from
two-electron collisions with an effective potential like that of
Eq.~(\ref{eq_potOv}).

We have evaluated the resulting pair distribution $g(r)$, not just for
$r=0$ [Fig.~\ref{fig_g0} and Eq.~(\ref{eq_fitg0}), which confirm 
the accuracy of the
Yasuhara\cite{Yasuhara} $g(0)$] but for all $r$ (Fig.~\ref{fig_totg}). We find
good agreement with Quantum Monte Carlo data in the range $0.5\, r_s
\lesssim r \lesssim r_s$, where this data is accurate, while our
results are probably more reliable than the Monte Carlo data
for $r\lesssim 0.5\,r_s$.

We have also studied the small-$r$ expansion of the spin-resolved
pair-distribution functions, Eqs.~(\ref{eq_expgud}) and~(\ref{eq_expguu}).
The $r_s$-dependent coefficients $a_2^{\sigma\sigma'}(r_s)$ of 
$(r/r_s)^2$ from the Overhauser model have been extracted and then
corrected for the known high-density limit\cite{Rassolov,reply} 
[Fig.~\ref{fig_a2}, Eqs.~(\ref{eq_a2udfit}) and~(\ref{eq_a2uufit})
with corrected $\beta_a$ and $d_p$].
Unlike $g(0)$ and $a_2^{\uu}$, our $a_2^{\ud}$ has a non-monotonic
dependence upon $r_s$. We have also discussed how to average the
short-range coefficients of $g(r)$ over coupling constant 
[Eqs.~(\ref{eq_a0av})-(\ref{eq_auuav})], and how to scale 
them for relative spin polarization
$\zeta \ne 0$ [Eqs.~(\ref{eq_a2zeta}) and~(\ref{eq_g0z})]. 
In future work, we plan to use
these results to make a new analytic model for the uniform-gas
$g(r)$ at all $r$, for use in density functional theory. Unlike
currently available models which fail at very small or very large
$r_s$, this new model will be designed to work over the whole range
$0<r_s<\infty$, while satisfying more exact constraints.

Except very close to $r=0$, highly accurate pair-distribution functions
for the uniform electron gas can be found from Quantum
Monte Carlo simulations, or from the fluctuation-dissipation
theorem using\cite{LGP} RPA-like response functions with the 
Richardson-Ashcroft\cite{RA} local-field factor. By supplementing
that approach with our small-$r$ expansions of Eqs.~(\ref{eq_expgud}) 
and~(\ref{eq_expguu}), it should be possible to find essentially-exact
spin-resolved pair-distribution functions and correlation energies.

Finally, we suspect that this approach can be applied usefully to other
systems, including the two-dimensional electron gas.\cite{VMS,polini}

\section*{Acknowledgments}
This work was supported in part by the
{\it Fondazione Angelo Della Riccia} (Firenze, Italy), and in part by 
the U.S. National Science Foundation under grant No. DMR98-10620.
We are very grateful to A.~W.~Overhauser for sending us the derivation
of Eq.~(\ref{eq_pk}) and for many useful comments at an early stage
of our work, and to Paul Ziesche for many helpful suggestions.

\end{document}